\documentclass[universe,article,accept,moreauthors,10pt,a4paper]{mdpi} 
\usepackage[english]{babel}
\usepackage{color} 

\firstpage{1} 
\articlenumber{x}
\doinum{10.3390/------}
\pubvolume{3}
\pubyear{2017}
\copyrightyear{2017}
\externaleditor{Academic Editors: Roman Pasechnik, Jos\'e Eliel
  Camargo-Molina and Ant\'onio Pestana Morais}
\history{Received: 29 October 2016; Accepted: 15 January 2017; Published: date}

\def\dblone{\hbox{$1\hskip -1.2pt\vrule depth 0pt height 1.6ex width 0.7pt
     \vrule depth 0pt height 0.3pt width 0.12em$}}

\newcommand{\okgr}{\Omega_{\rm K}}
 
\usepackage{booktabs} 
\usepackage{multirow}
\usepackage{soul} 
\usepackage{microtype}

\Title{Quark Deconfinement in Rotating Neutron Stars}

\Author{Richard D. Mellinger, Jr$^{1}$, Fridolin Weber
  $^{1,2,}$*, William Spinella $^{1}$, Gustavo A. Contrera $^{3,4,5}$
  \linebreak and Milva G. Orsaria $^{1,3,4}$}

\AuthorNames{Richard D. Mellinger Jr, William Spinella, Fridolin
  Weber, Gustavo A. Contrera, and Milva G. Orsaria}

\address {$^{1}$ \quad Department of Physics, San Diego State
  University,
  San Diego, CA 92182, USA;
  imasillypirate@gmail.com (R.D.M.); william.spinella@gmail.com
  (W.S.); morsaria@fcaglp.unlp.edu.ar (M.G.O.) \\
$^{2}$ \quad Department of Physics, University of California, San Diego, La Jolla, CA
  92093, USA \\
  $^{3}$ \quad National Scientific and Technical Research Council (CONICET),
  Godoy Cruz 2290, Buenos Aires (1425), Argentina; contrera@fisica.unlp.edu.ar \\
$^{4}$ \quad Grupo de Gravitaci\'{o}n, Astrof\'{\i}sica y
  Cosmolog\'{\i}a, Facultad de Ciencias Astron\'{o}micas y
  Geof\'{\i}sicas, Universidad Nacional de La Plata,  La Plata (1900),  Argentina \\
  $^{5}$ \quad Instituto de F\'{\i}sica La Plata,
National Scientific and Technical Research Council (CONICET), Universidad Nacional de La Plata, 
\mbox{La  Plata} (1900), Argentina} 

\corres{Correspondence: fweber@mail.sdsu.edu}

\abstract{In this paper, we use a three flavor non-local
  Nambu--Jona-Lasinio (NJL) model, an~improved effective model of
  Quantum Chromodynamics (QCD) at low energies, to investigate the
  existence of deconfined quarks in the cores of neutron stars.
  Particular emphasis is put on the possible existence of quark matter
  in the cores of rotating neutron stars (pulsars).  In contrast to
  non-rotating neutron stars, whose particle compositions do not
  change with time (are frozen in),
  the type and structure of the
  matter in the cores of rotating neutron stars depends on the spin
  frequencies of these stars, which opens up a possible new window on
  the nature of matter deep in the cores of neutron stars. Our study
  shows that, depending on mass and rotational frequency, up to around
  8\% of the mass of a massive neutron star may be in the mixed
  quark-hadron phase, if the phase transition is treated as a Gibbs
  transition. We also find that the gravitational mass at which quark
  deconfinement occurs in rotating neutron stars varies quadratically
  with spin frequency, which can be fitted by \mbox{a simple formula}.}


\begin{document}



\section{Introduction}

Exploring the properties of compressed baryonic matter, or, more
generally, strongly interacting matter at high densities and/or
temperatures has become a forefront area of modern physics
\cite{CBMbook11:a,NICA09:a,bass}. Experimentally, the properties of such
matter are being probed with the Relativistic Heavy Ion Collider RHIC
at Brookhaven and the Large Hadron Collider (LHC) at Cern. Great
advances in our understanding of such matter are expected from the
next generation of heavy-ion collision experiments at FAIR (Facility
for Antiproton and Ion Research, GSI), NICA (Nucloton-based Ion
Collider fAcility, JINR), and, at lower energies, from radioactive
beam facilities such as FRIB at MSU.

The universe was filled with hot and dense baryonic matter shortly
after the Big Bang. \mbox{Today, such matter} is being created in the
universe in the final stages of catastrophic stellar events (e.g.,
core-collapse supernovae, gamma-ray bursts) and exists permanently
inside of neutron stars.  Depending on mass and rotational frequency,
gravity compresses the matter in the core regions of neutron stars to
densities that are several times higher than the density of ordinary
atomic nuclei~\cite{glen97:book,weber99:book,blaschke01:trento,lattimer01:a,weber05:a,%
  page06:review,klahn06:a,sedrakian07:a,klahn07:a}. At such huge
densities atoms themselves collapse, and atomic nuclei are squeezed so
tightly together that new particle states may appear and novel states
of matter, foremost quark matter, may be created.  This feature makes
neutron stars superb astrophysical laboratories for \mbox{a wide range of}
physical studies~\cite{glen97:book,weber99:book,blaschke01:trento,%
  lattimer01:a,page06:review,sedrakian07:a,alford08:a,becker09:a}.  With
observational data accumulating rapidly from both orbiting and ground
based observatories spanning the spectrum from X-rays to radio
wavelengths, there has never been a more exiting time than today to
study neutron stars.  The Hubble Space Telescope and X-ray satellites
such as Chandra and XMM-Newton, for instance, have proven especially
valuable.  New astrophysical instruments such as the Five hundred
meter Aperture Spherical Telescope (FAST), the square kilometer Array
(skA), Fermi Gamma-ray Space Telescope (formerly GLAST), Astrosat,
ATHENA (Advanced Telescope for High ENergy Astrophysics), and the
Neutron Star Interior Composition Explorer (NICER) promise the
discovery of tens of thousands of new neutron stars. In particular,
the NICER mission is expected to open up a window on the inner
workings of neutron stars, as this mission is dedicated to the study
of extraordinary gravitational, electromagnetic, and nuclear physics
environments embodied by neutron stars.

In this paper, we use a non-local extension of the SU(3)
Nambu--Jona-Lasinio (NJL) model to investigate the possible existence
of deconfined quarks in the cores of neutron stars. Investigations of
dense neutron star matter based on the non-local SU(2) version of the
NJL model have been carried out in
Refs.\ \cite{Grigorian:2004,Blaschke:2007}.

The possible existence of such matter inside of neutron stars has
already been suggested in the 1970s \cite{fritzsch73:a} but has
remained an open issue ever since, as QCD cannot be solved for dense
astrophysical matter with the tools usually employed by relativistic
quantum field theories. We put particular emphasis on the possible
existence of quark matter inside of rotating neutron stars, known~as
pulsars.  In contrast to non-rotating neutron stars, whose core
compositions do not change with time \linebreak(are frozen in), the type and
structure of the matter in the cores of rotating neutron stars depends
on the spin frequencies of these stars, which could give rise to
observable astrophysical signals of quark deconfinement
\cite{glen97:a,chubarian00:a,glen01:clustering1,glen01:clustering2,poghosyan01:a}.

\begin{figure}[H]
\begin{center}
\includegraphics[scale=0.45]{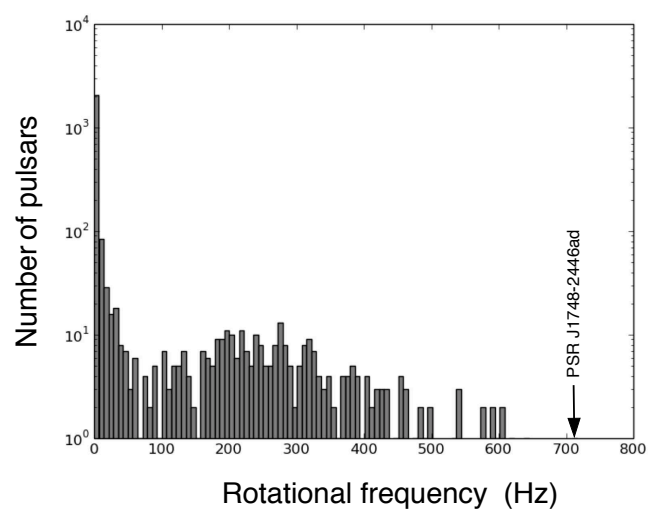}
\end{center}
\caption{Histogram for the frequencies of all 2510 pulsars which have
  frequency data in version 1.53 of The Australia Telescope National
  Facility Pulsar Catalog \cite{Manchester:2004bp,ATNF}. The most
  rapidly rotating neutron star observed to date is the pulsar
  J1748-2446ad, which rotates at a frequency of 716 Hz
  \cite{Hessels:2006ze} (rotational period of 1.4
  milliseconds).}
\label{FreqHist}
\end{figure}

A histogram of the rotational frequencies of 2510
observed pulsars is shown in Figure \ref{FreqHist}. T\mbox{his number} is
expected to increase dramatically once new instruments such as FAST
and skA come \mbox{into operation}.

The paper is organized as follows. In Section\ \ref{sec:eos} we
introduce the nuclear equations of state (EoS) that were used to model
the properties of (non) rotating neutron stars. The treatment of
rotating neutron stars in the framework of general relativity theory
is discussed in Section\ \ref{sec:GR}. The results of our investigations
are presented in Section\ \ref{Sec:Results}. A brief discussion of the
results in provided in Section\ \ref{Sec:Discussion}.

\section{Models for the Nuclear Equations of State \label{sec:eos}}
\vspace {-6pt}

\subsection{Hadronic Matter}

In the most primitive conception, the matter in the core of a neutron
star is constituted from neutrons. At a slightly more accurate
representation, the cores consist of neutrons and protons whose
electric charge is balanced by leptons
($\lambda=\{e^-,\mu^-\}$). Other particles, like hyperons \linebreak
($B=\{n,p,\Lambda,\Sigma,\Xi\}$) and the $\Delta$ resonance, may be
present if the Fermi energies of these particles become large enough
so that the existing baryon populations can be rearranged and a lower
energy state be reached.  To model this hadronic phase, we make use of
the relativistic mean-field (RMF) approximation, in which the
interactions between baryons are described by the exchange of scalar
($\sigma$), vector ($\omega$), and isovector ($\rho$) mesons
\cite{walecka74:a}.  The standard mean-field Lagrangian is given by
\cite{glen85:b,glen97:book,weber99:book,boguta77,boguta77:a,boguta83:a}.

{{\fontsize{9}{9}\selectfont  {\begin{eqnarray} 
&&\mathcal{L} = \sum_B   \bar\psi_B  \left[
    \left( i \gamma^{\mu} \partial_{\mu} - m_B \right) 
    + \Gamma_{\sigma B} -  \gamma_\mu \Gamma_{\omega B}^\mu  - \gamma_\mu
    \Gamma_{\rho B}^\mu 
    \boldsymbol{\tau} \right] \psi_B
    +\frac{1}{2}(\partial_{\mu}\sigma\partial^{\mu}\sigma-
  m^2_{\sigma}\sigma^2) -\frac{1}{3}b_{\sigma}m_n (g_{\sigma B}\sigma)^3
  \nonumber \\ &&- \frac{1}{4}c_{\sigma} (g_{\sigma B}\sigma)^4 -
  \frac{1}{4}\omega_{\mu\nu} \omega^{\mu\nu}
  +\frac{1}{2}m^2_{\omega}\omega_{\mu}\omega^{\mu} +\frac{1}{2}
  m^2_{\rho}\boldsymbol{\rho}_{\mu}\cdot \boldsymbol{\rho}^{\mu}
  -\frac{1}{4}\boldsymbol{\rho}_{\mu\nu}\cdot
  \boldsymbol{\rho}^{\mu\nu}
  +\sum\limits_{\lambda}\bar\psi_{\lambda}(i\gamma^{\mu}
  \partial_{\mu}-m_\lambda)\psi_{\lambda} \, , \label{eq:lagrangian}
\end{eqnarray}}}

with the meson-baryon vortices given by
\begin{eqnarray}
  \Gamma_{\sigma B} = g_{\sigma B} \, \sigma\, , \quad
  \Gamma_{\omega B}^\mu = g_{\omega B} \,  \omega^\mu \, , \quad
  \Gamma_{\rho B}^\mu = g_{\rho B} \, \boldsymbol{\rho}^\mu \, .
  \label{eq:RMFvortices}
\end{eqnarray}

The $\sigma$ and $\omega$ mesons in Equation\ (\ref{eq:lagrangian}) are
responsible for nuclear binding while the $\rho$ meson is required to
obtain the correct value for the empirical symmetry energy. The cubic
and quartic $\sigma$ terms in Equation\ (\ref{eq:lagrangian}) are
necessary, at the relativistic mean-field level, to obtain the
empirical incompressibility of nuclear matter
\cite{boguta77,boguta77:a}. The field tensors $\omega_{\mu\nu}$ and
${\boldsymbol\rho}_{\mu\nu}$ are defined as $\omega_{\mu\nu} =
\partial_\mu\omega_\nu - \partial_\nu\omega_\mu$ and
${\boldsymbol\rho}_{\mu\nu} = \partial_\mu{\boldsymbol\rho}_\nu -
\partial_\nu{\boldsymbol\rho}_\mu$.

The parameters (i.e., coupling constants) of the theory must reproduce
the bulk properties of infinite nuclear matter at saturation density,
$\rho_0=0.16~ {\rm fm}^{-3}$.  These are the binding energy $E/N$,
effective nucleon mass $m^*_N/m_N$, nuclear incompressibility $K$, and
the symmetry energy $a_{\rm sy}$ and its density derivative $L$. Of
the six, the values of $K$, $a_{\rm sy}$, and $L$ carry some
uncertainty.  The $K$ value is believed to lie in the range between
about 220 and 260~MeV \cite{shlomo06:a}, or between 250 and 315~MeV,
as recently suggested in Reference\ \cite{stone14:a}.  The values for
$a_{\rm sy}$ and $L$ are in the ranges of 29 to 35 MeV and 43 to 70
MeV, respectively \cite{krueger13:a,lattimer14:a,danielewicz:14a}.  We
have chosen two parameter sets, denoted GM1 and DD2, which~approximately cover the uncertainties in the nuclear matter properties
just above. They are listed in~Table \ref{tab:nmprop}.

The coupling constants associated with GM1 are $g_{\sigma B}= 9.572$, $g_{\omega
  B}= 10.618$, $g_{\rho B} =8.198$, $b_{\sigma} = 0.002936$, and~$c_{\sigma}= -0.00107$ \cite{glendenning91:a}. 

\begin{table}[H]
\centering
\begin{tabular}{c@{\hskip 1cm}c@{\hskip 1cm}c@{\hskip 1cm}c@{\hskip 1cm}}
  \toprule
\textbf{Nuclear Matter Property}   &\textbf{Units} &\textbf{GM1} &\textbf{DD2} \\
\midrule
$\rho_{0}$  &fm$^{-3}$  & 0.153      &0.149 \\
E/N         &MeV       & $-16.3$    &$-16.0$\\
K           &MeV       & 300        &243  \\
${m_{N}^{*}}/{m_{N}}$   &    & 0.70       &0.56  \\
$a_{sy}$    &MeV        & 32.5       &32.7\\
$L$        &MeV        &91.96       & 55.04 \\
\bottomrule
\end{tabular}
\caption{Properties of infinite nuclear matter at saturation density
  computed for parameter sets GM1 \cite{glendenning91:a} and DD2
  \cite{typel10:a}. Shown are the saturation density $\rho_{0}$,
  energy per nucleon $E/N$, nuclear incompressibility $K$, effective
  nucleon mass $m_N^*/m_N$, asymmetry energy $a_{sy}$, and the density
  derivative of the symmetry energy, $L$.}
\label{tab:nmprop}
\end{table}

The field equations for the baryon fields $\psi_B$, which follow from
Equation\ (\ref{eq:lagrangian}), are given by~\cite{weber05:a,glen85:b,glen97:book,weber99:book}.
\begin{eqnarray}
\left( i \gamma^\mu\partial_\mu-m_B \right) \psi_B = - g_{\sigma B}
\sigma \psi_B + g_{\omega B}\gamma^\mu\omega_\mu \psi_B + g_{\rho B}
\gamma^\mu \ {\boldsymbol\tau}\boldsymbol\cdot{\boldsymbol\rho}_\mu
\psi_B \, .
\label{eq:eompsi}
\end{eqnarray}

The meson fields in Equation (\ref{eq:eompsi}) are solutions of the following
field equations \cite{weber05:a,glen85:b,glen97:book,weber99:book}.
\begin{eqnarray}
\left( \partial^\mu\partial_\mu+m^2_\sigma \right) \sigma &=& \sum_B
g_{\sigma B} \bar\psi_B \psi_B - m_N b_N g_{\sigma N} \left( g_{\sigma
  N} \sigma \right)^2 - c_N\, g_{\sigma N} \left(
g_{\sigma N} \sigma \right)^3 \, ,
\label{eq:eomsigma} \\
\partial^\mu \omega_{\mu\nu} + m_\omega^2 \, \omega_\nu &=& \sum_B
g_{\omega B} \bar\psi_B \gamma_\nu \psi_B \, ,
\label{eq:eomom5} \\
\partial^\mu {\boldsymbol\rho}_{\mu\nu} + m_\rho^2 \,
\boldsymbol\rho_\nu &=& \sum_B  g_{\rho B} \bar\psi_B
\boldsymbol\tau \gamma_\nu \psi_B \, .
\label{eq:eomrho5}
\end{eqnarray}

In the standard RMF limit, the meson fields of Equations (\ref{eq:eomsigma})--(\ref{eq:eomrho5}) simplify to
\cite{weber05:a,glen85:b,glen97:book,weber99:book}.
\begin{eqnarray} 
    m^2_{\sigma}\bar\sigma &=& \sum_B g_{\sigma
      B}\frac{2J_B+1}{2\pi^2} \int_0^{k_B}
    \frac{m_B^*(\sigma)}{\sqrt{k^2+m_B^{*2}(\sigma)}}\,k^2 dk -b m_N
    g_{\sigma} (g_{\sigma} {\bar\sigma})^2 -c g_{\sigma} (g_{\sigma}
    {\bar\sigma})^3 \, ,
    \label{eq:sigma} \\
  \bar\omega_0 &=& \sum_B \frac{g_{\omega B}}{m_{\omega}^2}\rho_B \, ,
 \label{eq:omega} \\
 \bar\rho_{03} &=& \sum_B \frac{g_{\rho B}}{m_{\rho}^2}I_{3B}\rho_B \, ,
 \label{eq:rho}
\end{eqnarray}
where $\bar\sigma$, $\bar\omega$, and $\bar\rho_{03}$ denote the
mean-field limits of $\sigma$, $\omega$, and $\bar{\boldsymbol\rho}$,
respectively, and the effective baryon masses are given by
$m^*_B(\bar\sigma) = m_B - g_{\sigma B}\bar\sigma$. $J_B$ and $I_{3B}$
denote the spin respectively isospin of a baryon of type $B$, and
$\rho_B$ stands for the number density of baryon $B$.

In addition to the standard RMF theory discussed just above, we also
consider a lagrangian where the meson-baryon vortices $\Gamma_{MB}$
(where $M = \sigma, \omega, \rho$) of Equation\ (\ref{eq:RMFvortices}) are
no longer constant but rather depend on density
\cite{brockmann92:a,lenske95:a,hofmann01:a}. In that case the values
of the vortices are derived from relativistic Dirac-Brueckner
calculations of nuclear matter, which use one-boson-exchange
interactions as an~input. A characteristic feature of the
density-dependent theory is the occurrence of rearrangement terms in
the expression for the baryon chemical potential, which leads to a
more complex condition for chemical equilibrium compared to the
standard RMF approximation \cite{hofmann01:a}.  The parameter set of
the density-dependent (DD) treatment adopted in this paper is denoted
DD2~($G_V=0$), where a~vanishing vector coupling constant $G_V=0$
among quarks has been chosen \cite{typel10:a}. For DD2, the coupling
constants at saturation density are $g_{\sigma N} = 10.687$,
$g_{\omega N} = 13.342$, and $g_{\rho N} = 3.627$ \cite{typel10:a},
which lead to the saturation properties of infinite nuclear matter
shown in Table \ref{tab:nmprop}.

The equation of state of the standard mean-field treatment (for the DD
formalism, see References~\cite{lenske95:a,hofmann01:a,fuchs95:a,typel99:a}) is
obtained by solving Equations\ (\ref{eq:sigma})--(\ref{eq:rho}) together
with the charge conservation conditions (baryonic, electric) given by
\cite{glen85:b,glen97:book,weber99:book}.
\begin{eqnarray}
\rho_b-\sum\limits_B \rho_B=0 \, , \qquad  \sum\limits_{B} \rho_B
q_B +\sum\limits_{\lambda} \rho_{\lambda} q_{\lambda}=0 \, ,
\end{eqnarray}
where $\rho_b$ is the total baryonic density and $q_B$ and $q_\lambda$
are the electric charges of baryons and leptons, respectively.
Particles in the hadronic phase are subject to the chemical
equilibrium condition
\begin{equation} \label{eq:chemicalequilibrium}
  \mu_B = b_B \mu_n - q_B \mu_e \, ,
\end{equation}
where $\mu_B$ is the chemical potential and $b_B$ is the baryon number
of baryon $B$. $\mu_n$ and $\mu_e$ denote the linearly independent
chemical potentials of neutrons and electrons, respectively, which
reflect baryon number and electric charge conservation on neutron star
matter.  New baryon or lepton states are populated when the right side
of Equation (\ref{eq:chemicalequilibrium}) is greater than the
particle's chemical potential. The~baryonic and leptonic number
densities ($\rho_B$ and $\rho_{\lambda}$) are both given by $\rho_i =
(2J_i+1) k_i^3 / 6\pi^2$.  The~unknowns of the theory are the meson
mean-fields ($\sigma$, $\omega$, $\rho$), and the neutron and electron
fermi momenta ($k_n$, $k_e$). Finally, the energy density and pressure
of the hadronic phase are given by
\cite{glen85:b,glen97:book,weber99:book}.

{{\fontsize{9}{9}\selectfont  {\begin{eqnarray} 
    \epsilon_H &=&\frac{1}{3} b m_N (g_{\sigma N} \bar\sigma)^3 +\frac{1}{4}
    c (g_{\sigma N} \bar\sigma)^4 +\frac{1}{2} m_{\sigma}^2 \bar\sigma^2
    +\frac{1}{2} m_{\omega}^2 \bar\omega_0^2
    +\frac{1}{2} m_{\rho}^2 \bar\rho_{03}^2 + \sum_B \frac{2J_B+1}{2\pi^2}
    \int_0^{k_B}\sqrt{k^2+m_B^{*2}} \, k^2 dk \nonumber \\ &&+
  \frac{1}{\pi^2} \sum_{\lambda}  \int_0^{k_{\lambda}}
    \sqrt{k^2+m_{\lambda}^2} \, k^2 dk \, , \label{eq:hdensity} \\
    p_H &=& - \frac{1}{3} b m_N (g_{\sigma N} \bar\sigma)^3
    -\frac{1}{4} c (g_{\sigma N} \bar\sigma)^4 -\frac{1}{2} m_{\sigma}^2 \bar\sigma^2
    +\frac{1}{2} m_{\omega}^2 \bar\omega_0^2
    +\frac{1}{2} m_{\rho}^2 \bar\rho_{03}^2 + \frac{1}{3}\sum_B
    \frac{2J_B+1}{2\pi^2} \int_0^{k_B}\frac{k^4
      dk}{\sqrt{k^2+m_B^{*2}}} \nonumber
    \\ &&+\frac{1}{3}\sum_{\lambda} \frac{1}{\pi^2}
    \int_0^{k_{\lambda}} \frac{k^4 dk}{\sqrt{k^2+m_{\lambda}^2}} \,
    . \label{eq:hpressure}
\end{eqnarray}}}

\subsection{Deconfined Quark Phase}\label{ss:quarkphase}

A popular model widely used to describe deconfined 3-flavor (up, down,
strange) quark matter is the Nambu--Jona-Lasinio model
\cite{NJL61:a,NJL61:b,Klevansky:1992,vogl:1995,buballa05:a}.  Here we
use a non-local extension of this model \linebreak(n3NJL) \cite{orsaria13:a,orsaria14:a}, whose effective action is \mbox{given by}
\begin{eqnarray}
S_E &=& \int d^4x \left\{ \bar \psi (x) \left[ -i
  \partial{\hskip-2.0mm}/ + \hat m \right] \psi(x) - \frac{G_S}{2}
\left[ j_a^S(x) \ j_a^S(x) + j_a^P(x) \ j_a^P(x) \right] \right.
\nonumber \\ & & \qquad \qquad \left. - \frac{H}{4} \ T_{abc} \left[
  j_a^S(x) j_b^S(x) j_c^S(x) - 3\ j_a^S(x) j_b^P(x) j_c^P(x)
  \right]\right.  \left.- \frac{G_{V}}{2} \left[j_{V}^\mu(x)
  j_{V}^\mu(x)\right]\right\}\, , \label{L3}
\end{eqnarray}
where $\psi \equiv (u, d, s)^T$, $\hat m = {\rm diag}(m_u, m_d, m_s)$
is the current quark mass matrix, $\lambda_a$ ($a=1,...,8$) denote the
Gell-Mann matrices--generators of SU(3), and $\lambda_0=\sqrt{2/3}\,
\dblone_{3\times 3}$.  The currents $j_a^{S,P}(x)$ and
$j_{V}^{\mu}(x)$ are given by
\begin{align}
j_{a}^S(x) & =\int d^{4}z\ \widetilde{g}(z)\ \bar{\psi}\left(
x+\frac{z}{2}\right) \ \lambda_{a}\ \psi\left( x-\frac{z}{2}\right)
\ ,\nonumber\\ j_{a}^P(x) & =\int
d^{4}z\ \widetilde{g}(z)\ \bar{\psi}\left( x+\frac{z}{2}\right) \ i
\ \gamma_5 \lambda_{a} \ \psi\left(
x-\frac{z}{2}\right)\ ,\nonumber\\ j_{V}^{\mu}(x) & =\int
d^{4}z\ \widetilde{g}(z)\ \bar{\psi}\left( x+\frac{z}{2}\right)
\ \gamma^{\mu}\lambda_{a}\ \psi\left( x-\frac{z}{2}\right), \label{currents}
\end{align}
where $\widetilde{g}(z)$ is a form factor responsible for the
non-local character of the interaction. Finally,~the~constants
$T_{abc}$ in the t'Hooft term accounting for flavor-mixing are defined
by
\begin{equation}
T_{abc} = \frac{1}{3!} \ \epsilon_{ijk} \ \epsilon_{mnl} \
\left(\lambda_a\right)_{im} \left(\lambda_b\right)_{jn}
\left(\lambda_c\right)_{kl}\;.
\end{equation}

The current quark mass $\bar{m}$ of up and down quarks and the
coupling constants $G_S$ and $H$ in Equation\ (\ref{L3}), have been
fitted to the pion decay constant, $f_\pi$, and meson masses
$m_{\pi}$, $m_\eta$, and~$m_{\eta'}$, as described in
\citep{Contrera2008,Contrera2010}. The~result of this fit is $\bar{m}
= 6.2$~MeV, $\Lambda = 706.0$~MeV, $G_S \Lambda^2 = 15.04$, $H
\Lambda^5 = - 337.71$. The strange quark current mass is treated as a
free parameter and was set to $m_s =140.7$~MeV.  The strength of the
vector interaction $G_V$ is expressed in terms of the strong coupling
constant $G_S$. To account for the uncertainty in the theoretical
predictions for the ratio $G_V/G_S$, we treat the vector coupling
constant as a free parameter
\cite{Sasaki:2006ws,Fukushima:2008wg,Bratovic:2012qs}, which varies
from $0$ to $0.09 \, G_S$.

For the mean-field approximation, the thermodynamic potential
following from $S_E$ of Equation~\ (\ref{L3}) is given~by
\begin{eqnarray} 
    &&\Omega = -\frac{3}{\pi^3}\sum_{f=u,d,s}
    \int_0^{\infty}dp_0 \int_0^{\infty} dp \ \mathrm{ln}
    \Biggl\{ \left[ \widehat{\omega}^2_f+M_f^2(\omega^2_f)\right]
    \frac{1}{\omega^2_f+m^2_f}\Biggr\} 
    -\frac{3}{\pi^2}\sum_{f=u,d,s}\int_0^{\sqrt{\mu^2_f-m^2_f}}
    dp\,p^2 \times \nonumber \\
    &&\left[\left(\mu_f-E_f\right)\theta\left(\mu_f-m_f\right)\right]
    \quad -\frac{1}{2} \Bigg[\sum_{f=u,d,s}\left(\bar{\sigma}_f\bar{S}_f+
    \frac{G_S}{2}\bar{S}^2_f\right)+\frac{H}{2}\bar{S}_u\;\bar{S}_d\;\bar{S}_s
    \Biggr] 
   - \sum_{f=u,d,s}\frac{\overline{\omega}_f^2}{4G_V} \, , \label{grandpotential}
\end{eqnarray}
where $\bar{\sigma}_f$, $\overline{\omega}_f$, and $\bar{S}_f$ are the
quark scalar, vector, and auxiliary mean fields, respectively.
Moreover,~$E_f$~is given by $E_f= \sqrt{{\boldsymbol p}^2 + m^2_f}$
and we have defined $\omega^2_f = (p_0+i\mu_f)^2 + {\boldsymbol
  p}^2$. The~momentum dependent quark masses are given by
$M_f(\omega_f^2) = m_f+\bar{\sigma}_f g(\omega^2_f)$.  The quantities
$g(\omega^2_f) = \mathrm{exp}(-\omega^2_f/\Lambda^2)$ are the Gaussian
form factors, responsible for the nonlocal nature of the interaction
among quarks \cite{orsaria14:a}.  The~auxiliary mean fields are given
by
\begin{equation}
  \bar{S}_f = -48 \int_0^{\infty} dp_0 \int_0^{\infty}
  \frac{dp}{8\pi^3}g(\omega^2_f)\frac{M_f(\omega^2_f)}
  {\widehat{\omega}^2+M_f^2(\omega^2_f)} \, .
\end{equation}

The vector interactions taken into account in the treatment shift the
quark chemical potentials as \mbox{$\widehat{\mu}_f = \mu_f - g(w_f^2)
\overline{\omega}_f$} and $\widehat{\omega}_f^2 = (p_0 + i
\widehat{\mu}_f)^2+p^2$.  The scalar and vector mean fields are
obtained by minimizing the grand thermodynamic potential, $\partial
\Omega / \partial \bar{\sigma}_f = 0$ and $\partial \Omega / \partial
\overline{\omega}_f = 0$. Finally, the quark number densities are
obtained from $\rho_f = \partial \Omega / \partial \mu_f$.

To determine the equation of state one must solve a nonlinear system
of equations for the fields $\bar{\sigma}_f$ and $\bar{\omega}_f$, and
the neutron and electron chemical potentials $\mu_n$ and $\mu_e$.
This system of equations consists of the mean field equations
\begin{equation} \label{eq:q1}
  \bar{\sigma}_i + G_S \bar{S}_i + H\bar{S}_j \bar{S}_k = 0 \, ,
\end{equation}
with cyclic permutations over the quark flavors, $\overline{\omega}_f
-2 G_V \partial \Omega / \partial \overline{\omega}_f = 0$, and the
baryonic and electric charge conservation equations
$\sum_{f=u,d,s} \; \rho_f - 3 \rho_b = 0$ and $\sum_{f=u,d,s}\; \rho_f
\, q_f + \sum_{\lambda=e^-,\mu^-} \; \rho_{\lambda}\, q_{\lambda} =
0$, respectively.  Finally, the pressure $p_Q$ and
energy density $\epsilon_Q$ are given by
\begin{equation} \label{eq:qpressure}
  p_Q = \Omega_0-\Omega \, ,
\end{equation}
and
\begin{equation} \label{eq:qdensity}
  \epsilon_Q = -p_Q + \sum_{f=u,d,s} \rho_f \mu_f +
  \sum_{\lambda=e^-,\mu^-} \rho_{\lambda} \mu_{\lambda} \, ,
\end{equation}
where $\Omega_0$ was chosen by the condition that $P_Q$
vanishes at $T=\mu=0$ \cite{orsaria13:a,orsaria14:a}.

\subsection{Quark-Hadron Mixed Phase}

To model the mixed phase region of quarks and hadrons in neutron
stars, we use the Gibbs condition for phase equilibrium between
hadronic ($H$) and quark ($Q$) matter,
\begin{eqnarray}
  P_H ( \mu_n , \mu_e, \{ \phi \} ) = P_Q (\mu_n , \mu_e) \, ,
\label{eq:gibbs}
\end{eqnarray}
where $P_H$ and $P_Q$ denote the pressures of hadronic matter and
quark matter, respectively \cite{glen91:pt,glen01:b}. The~quantity
$\{ \phi \}$ in Equation\ (\ref{eq:gibbs}) stands collectively for the
field variables ($\bar{\sigma}$, $\bar\omega$, $\bar\rho$) and Fermi
momenta ($k_B$, $k_\lambda$) that characterize a~solution to the
equations of confined hadronic matter (Section\ \ref{sec:eos}). We use
the symbol $\chi \equiv V_Q/V$ to denote the volume proportion of
quark matter, $V_Q$, in the unknown volume $V$. By definition, $\chi$
varies between 0 and 1, depending on how much confined hadronic
matter has been converted to quark matter.  Equation (\ref{eq:gibbs})
is to be supplemented with the conditions of global baryon charge
conservation and global electric charge conservation.  The global
conservation of baryon charge is expressed as
\cite{glen91:pt,glen01:b}.
\begin{eqnarray}
  \rho_b = \chi \, \rho_Q(\mu_n, \mu_e ) + (1-\chi) \,
  \rho_H (\mu_n, \mu_e,  \{ \phi \}) \, ,
\label{eq:mixed_rho}
\end{eqnarray}
where $\rho_Q$ and $\rho_H$ denote the baryon number densities of the
quark phase and hadronic phase, respectively. The global neutrality of
electric charge is given by \cite{glen91:pt,glen01:b}.
\begin{eqnarray}
  0 = \chi \ q_Q(\mu_n, \mu_e ) + (1-\chi) \ q_H (\mu_n, \mu_e, \{
  \phi \}) \, ,
\label{eq:mixed_charge}
\end{eqnarray}
with $q_Q$ and $q_H$ denoting the electric charge densities of the
quark phase and hadronic phase, respectively.  We~have chosen global
rather than local electric charge neutrality.  Local NJL studies
carried out for local electric charge neutrality have been reported
recently in References\ \cite{bonanno12:a,masuda12:a,masuda13:a,lenzi12:a}.

In Figure\ \ref{fig:eos}, we show the GM1 and DD2 equations of state
(EoS) used in this work to study the quark-hadron composition of
rotating neutron stars.  The solid dots mark the beginning and the end
of the quark-hadron mixed phases for these equations of state.  Since
the Gibbs condition is used to model the quark-hadron phase
transition, pressure varies monotonically with the proportion of the
phases in equilibrium. This would not be the case if the Maxwell
construction were used to model the phase equilibrium between quarks
and hadrons, in which case the pressure throughout the mixed phase is
constant.  For that reason, the mixed phase would strictly be excluded
from neutron stars, but small cores made entirely of quark matter may
still be possible in neutron stars close the maximum-mass model (see,
for instance, Refrences\ \cite{sandoval16:a,contrera17:a}, and
references therein).

A Maxwell-like phase transition is generally supported by larger
surface tensions, $\sigma$, of quark matter (see, however,
Reference\ \cite{voskresensky02:a}). Values of $\sigma \sim 30\, {\rm
  MeV/fm}^2$ have recently been suggested in the literature
\cite{yasutake2014,surfacetension1,surfacetension2,surfacetension3,spinella16:a},
but its actual value is an open issue.

The GM1 and DD2 equations of state are compared in
Figure\ \ref{fig:eos} with models for the equation of state that have
recently been suggested in the literature. 'HLPS' and 'Neutron matter'
show the constraints on the equation of state established by Hebeler,
Lattimer, Pethick, and
Schwenk~\cite{krueger13:a,hebeler13:a}. \linebreak The curves labeled
'EoS I', 'EoS II', and 'EoS III' show the compact star matter
equations of state determined by Kurkela {et al.}\ \cite{kurkela14:a},
which are based on an interpolation between the regimes of low-energy
chiral effective theory and high-density perturbative QCD.  One sees
that the GM1 and DD2 models are well within these limits. The~only
difference concerns the behavior of the equation of state at
sub-nuclear densities (labeled ``Neutron matter'' in
Figure\ \ref{fig:eos}), where our models provide slightly more
pressure. This, however, does in no way impact the results for the
quark-hadron compositions shown in Sect.\ \ref{Sec:Results},
 because of the large masses
of these stars.

Figure \ref{fig:MecMR} shows the gravitational mass versus central
neutron stars density (left panel) and gravitational mass versus
radius of non-rotating neutron stars for the equations of state
discussed in this section. The maximum masses of these neutron stars
are between 1.9 and $2.1\, M_\odot$.  Stellar~rotation, which will be
discussed in the next section, increases the masses of these stars to
values that are between 2.2 and $2.4\, M_\odot$, while their central
densities drops by around 20\%.
\begin{figure}[tb]
\begin{center}
    \includegraphics[scale=0.4]{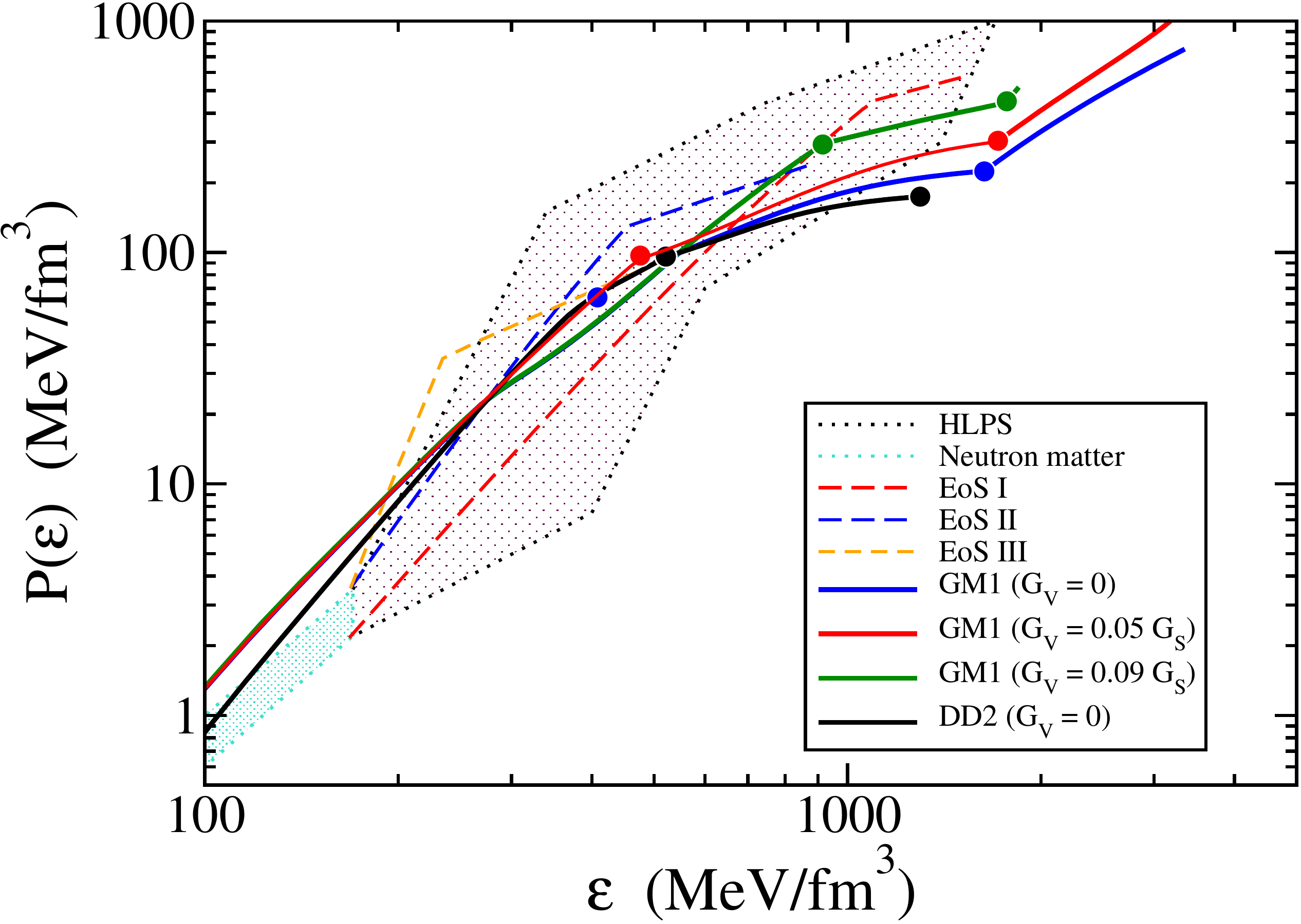}
\end{center}
\caption{Comparison of the equations of state used in this work, GM1
  and DD2, with models recently suggested in the literature (see text
  for details). The solid dots mark the beginning and the end of the
  quark-hadron mixed phases for GM1 and DD2. The repulsive interaction
  among quarks is controlled by the vector coupling constant, $G_V$.}
  \label{fig:eos}
\end{figure}
\begin{figure}[tb]
\begin{center}
  \includegraphics[scale=0.30]{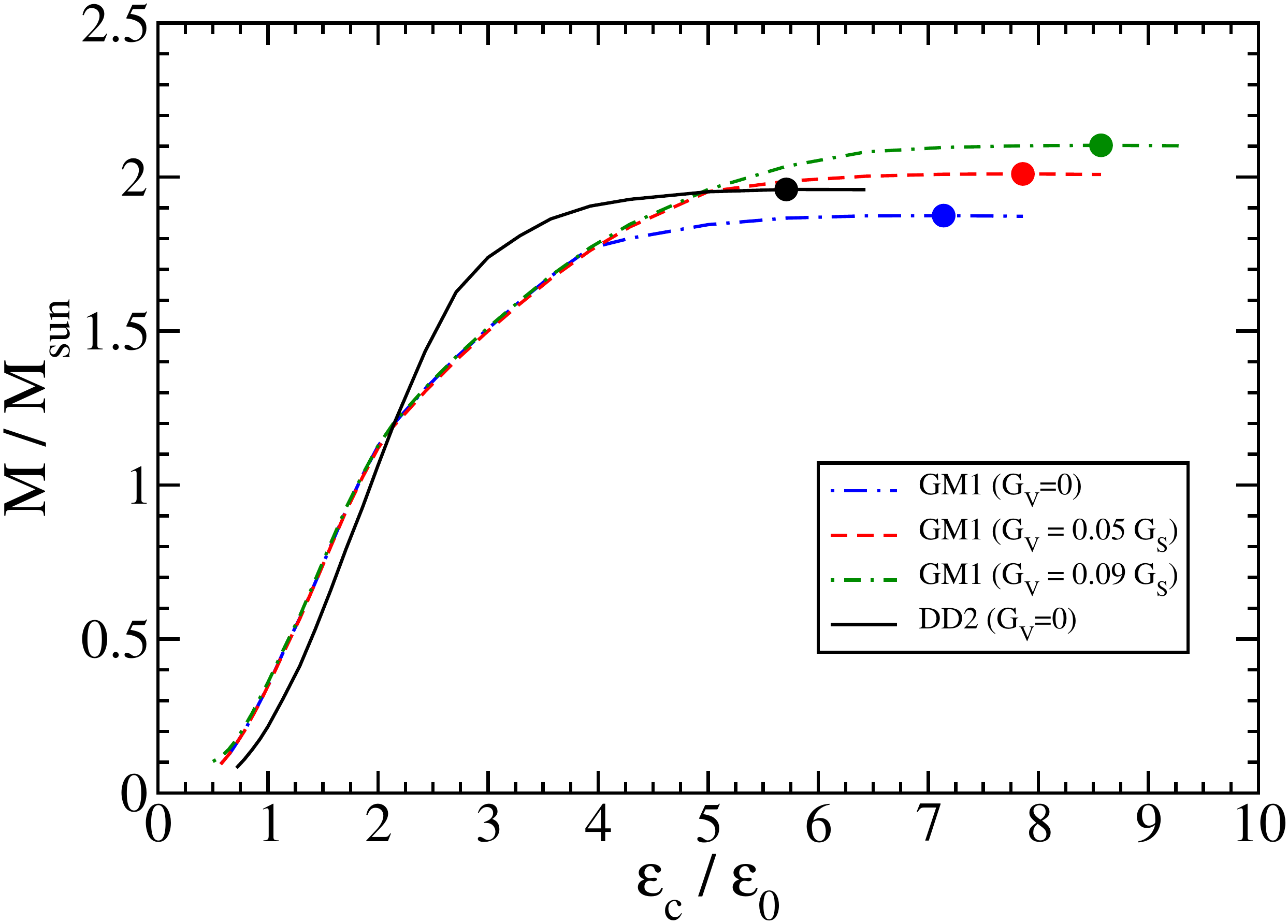}
  \includegraphics[scale=0.30]{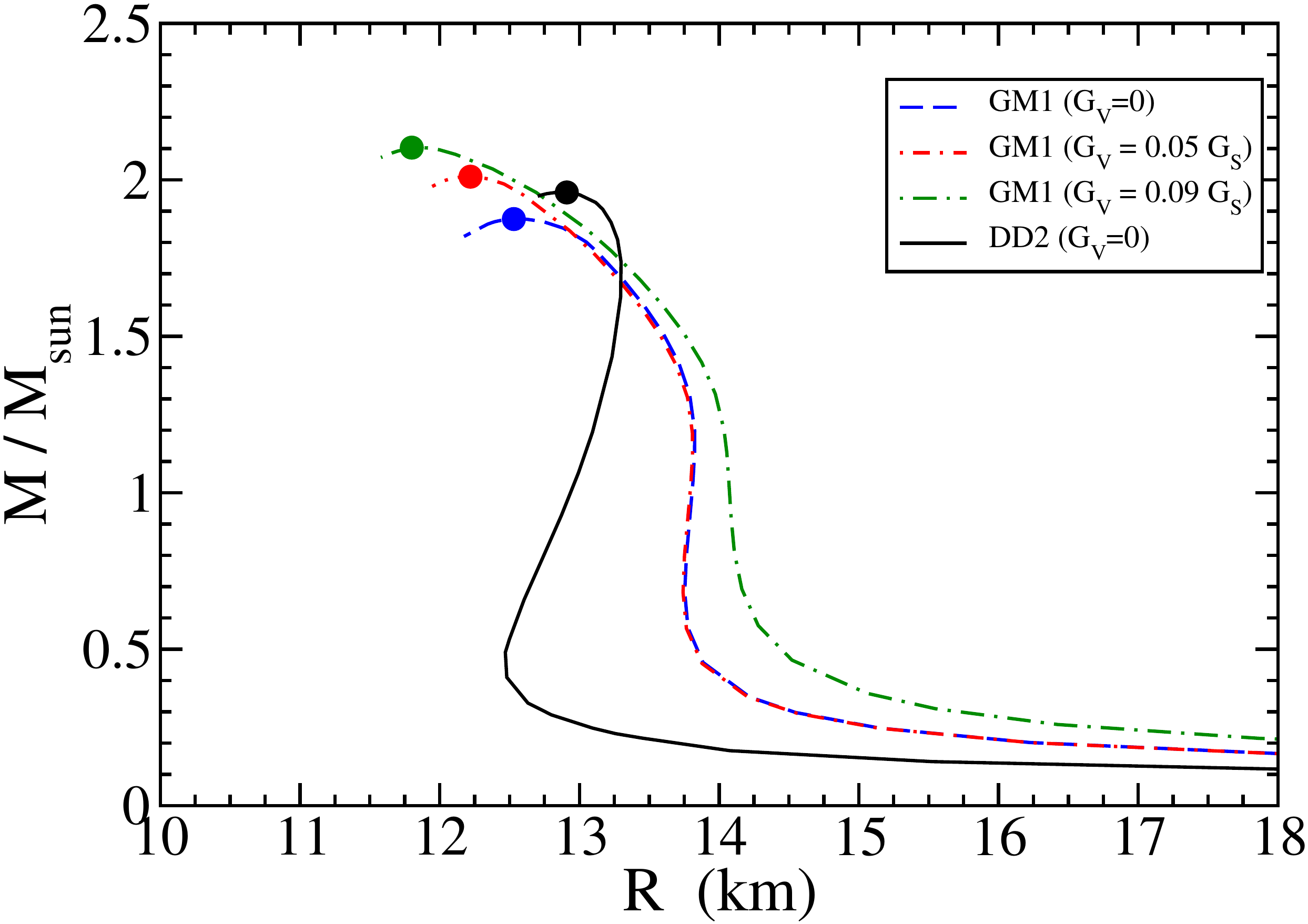}
\end{center}
\caption{Mass--central density (left) and mass--radius relationships
  of non-rotating neutron stars for the nuclear equations of state (EoS) used in this
  work. ($\epsilon_0 = 140$ MeV/fm$^3$ denotes the density of infinite
  nuclear matter).}
  \label{fig:MecMR}
\end{figure}

\section{Treatment of Rotating Neutron Stars in General Relativity
  Theory}\label{sec:GR}

The fact that rotation deforms neutron stars, stabilizes them against
collapse, and drags along the local inertial frames inside and outside
of them so that they co-rotate with the stars, renders the
construction of models of rotating neutron stars very complicated. A
line element which accounts for these three features has the form~\cite{weber99:book,friedman86:a,lattimer90:a,salgado94:a,cook94:a,%
  cook94:b}
\begin{eqnarray}
  d s^2 = - e^{2\nu} dt^2 + e^{2\psi} (d\phi - \omega dt)^2 + e^{2\mu}
  d\theta^2 + e^{2\lambda} dr^2 \, ,
\label{eq:metric} 
\end{eqnarray} where   $\nu$, $\psi$,  $\mu$ and
$\lambda$ denote metric functions and $\omega$ is the angular velocity
of the local inertial frames. All these quantities depend on the
radial coordinate $r$, the polar angle $\theta$, and implicitly on the
star's angular velocity $\Omega$.  The metric functions and the frame
dragging frequencies are to be computed from Einstein's field
equation,
\begin{equation} 
R^{\kappa\sigma} - \frac{1}{2} R g^{\kappa\sigma} = 8 \pi
T^{\kappa\sigma} \, ,
\label{eq:einstein}
\end{equation}
where $T^{\kappa\sigma} = T^{\kappa\sigma}(\epsilon, P(\epsilon))$
denotes the energy momentum tensor of the stellar matter, whose
equation of state is given by $P(\epsilon)$. The other quantities in
Equation\ (\ref{eq:einstein}) are the Ricci tensor $R^{\kappa\sigma}$, the
curvature scalar $R$, and the metric tensor, $g^{\kappa\sigma}$. No simple stability criteria are known for rapidly rotating stellar
configurations in general relativity. However, an absolute limit on
rapid rotation is set by the onset of mass shedding from the equator
of a rotating star. The corresponding rotational frequency is known as
the Kepler frequency, ${\Omega_{\,\rm K}}$. In classical mechanics,
the expression for the Kepler frequency, determined by the equality
between the centrifugal force and gravity, is readily obtained as
${\Omega_{\,\rm K}} = \sqrt{M/R^3}$. Its~general relativistic
counterpart is given by \cite{weber99:book,friedman86:a}.
\begin{eqnarray}
  {\Omega_{\,\rm K}} = \omega +\frac{\omega_{,r}} {2\psi_{,r}} +
  e^{\nu -\psi} \sqrt{ \frac{\nu_{,r}} {\psi_{,r}} +
    \Bigl(\frac{\omega_{,r}}{2 \psi_{,r}} e^{\psi-\nu}\Bigr)^2 } \, ,
\label{eq:okgr}  
\end{eqnarray} 
which is to be evaluated self-consistently at the equator of a
rotating neutron star. The Kepler period follows from
Equation\ (\ref{eq:okgr}) as ${P_{\,\rm K}} = {{2 \pi} / {\Omega_{\,\rm
      K}}}$. For typical neutron star matter equations of state, the~Kepler period obtained for $1.4\, M_\odot$ neutron stars is typically
around 1~ms.
\cite{weber99:book,weber05:a,friedman86:a,lattimer90:a}. An exception
to this are strange quark matter stars. Since they are self-bound,
they tend to possess smaller radii than conventional neutron stars,
which are bound by gravity only. Because of their smaller radii,
strange~stars can withstand mass shedding from the equator down to
periods of around 0.5~ms \cite{glen92:crust,glen92:limit}.

A mass increase of up to $\sim20$\% is typical for rotation at
${\Omega_{\,\rm K}}$. Because of rotation, the equatorial radii
increase by several kilometers, while the polar radii become smaller
by several kilometers. The~ratio between both radii is around 2/3,
except for rotation close to the Kepler frequency. The~most rapidly
rotating, currently known neutron star is pulsar PSR J1748--2446ad,
which rotates at a period of 1.39~ms (719~Hz) \cite{Hessels:2006ze}, well
below the Kepler frequency for most neutron star equations of state. Examples of other rapidly rotating neutron stars are PSRs
B1937+21 \cite{backer82:a} and B1957+20  \cite{fruchter88:a}, whose~rotational periods are 1.58 ms (633~Hz) and 1.61~ms (621~Hz),
respectively.

The density change in the core of a neutron star whose frequency
varies from $0 \leq \Omega \leq {\Omega_{\,\rm K}}$ can be as large as
60\% \cite{weber99:book,weber05:a}. This suggests that rotation may
drive phase transitions or cause significant compositional changes of
the matter in the cores of neutron stars
\cite{glen97:book,weber99:book,weber05:a}.

\section{Results}\label{Sec:Results}

For each equation of state, a numerical code, based on Hartle's
rotation formalism \cite{hartle67:a,hartle68:a,weber91:b}, was~used to
predict the percentage of the mass made up by deconfined quark matter
over a wide
range of frequencies and gravitational masses. Figures
\ref{fig:profiles} and \ref{fig:comp1} show the composition of
rotating neutron stars based on the lagrangian of
Equation\ (\ref{eq:lagrangian}). As can be seen, the quark-hadron mixed phase as well as several different hyperon species are successively
spun out of the neutron star if the rotation rate increases toward the
Kepler frequency.  Non-rotating neutron stars posses the most complex
compositions, since they are the most dense members of the rotational
sequence. The~compositions shown in Figure\ \ref{fig:comp1} are
snapshots taken from movies (see ''Supplementary Materials'' 
at the end of
the paper) showing the entire rotational evolution of this neutron
star all the way from zero frequency~to~$\Omega_{\rm K}$.

\begin{figure}[H]
\begin{center}
  \includegraphics[scale=0.27]{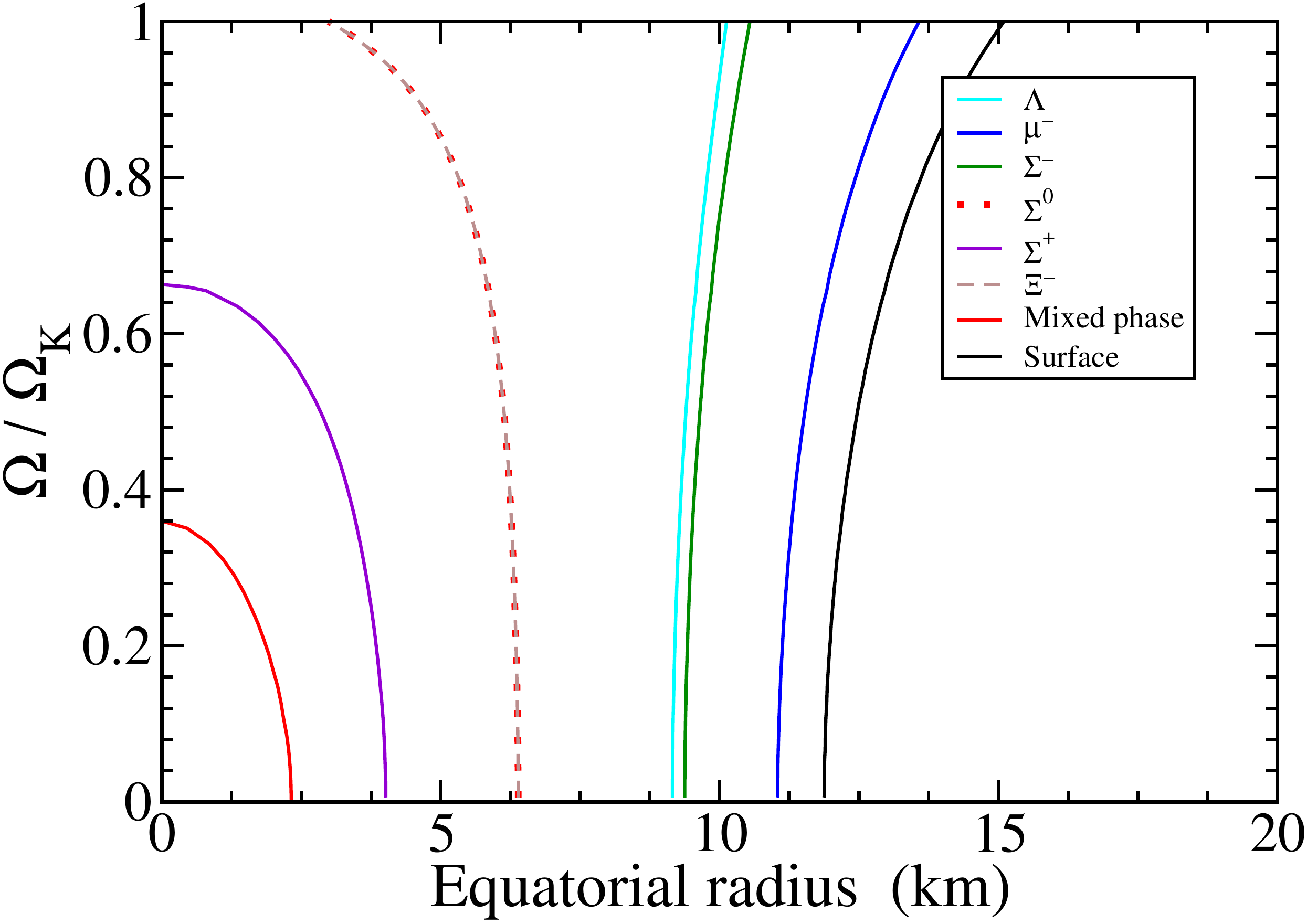}
  \includegraphics[scale=0.27]{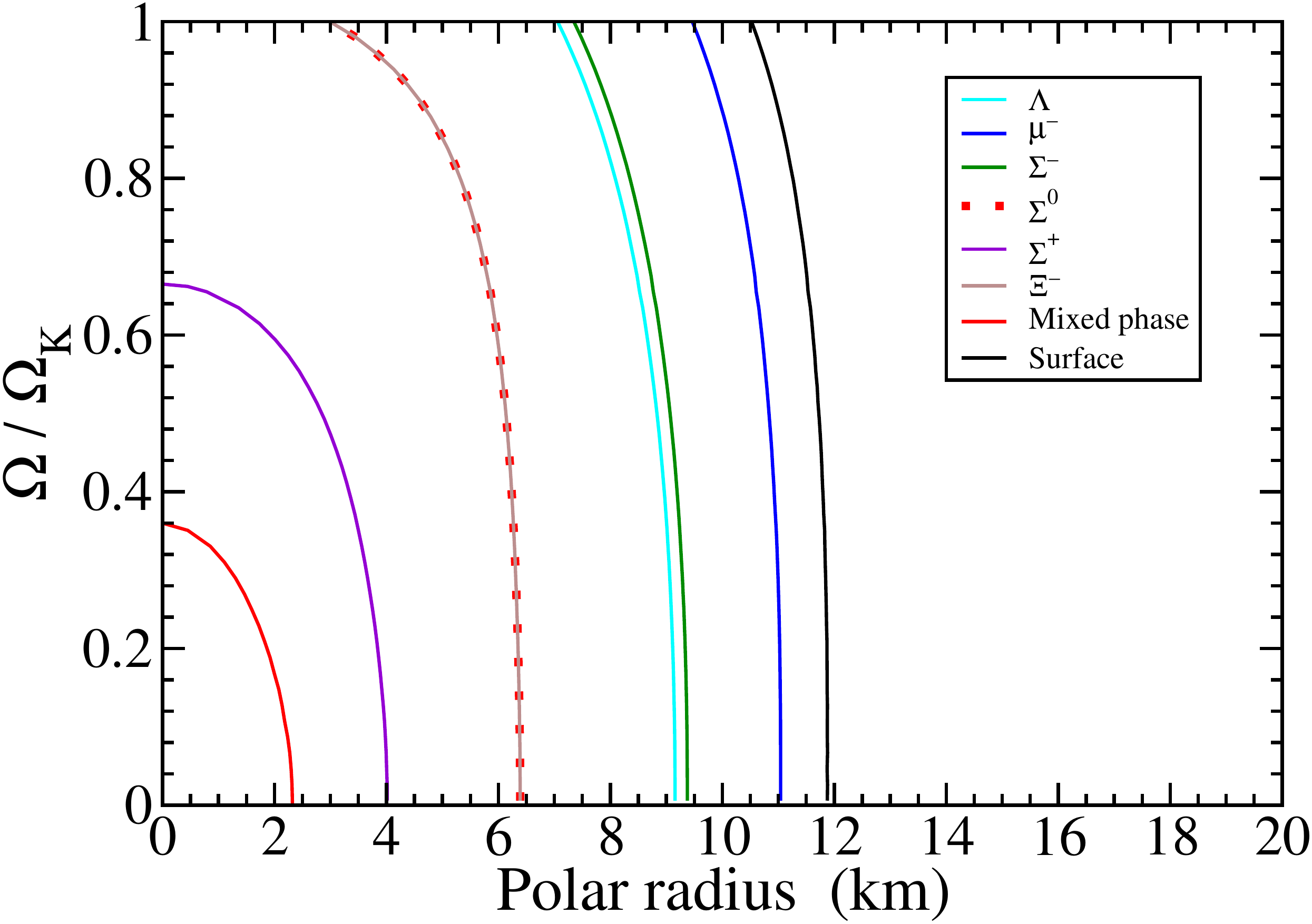}
\end{center}
\caption{Particle populations inside of rotating neutron stars, in
  equatorial (\textbf{left}) and polar (\textbf{right}) directions, computed for the GM1
  EoS. The vector interaction among quarks is $G_V=0.09 \, G_S$. The~  stellar frequency, $\Omega$, ranges from zero to the Kepler
  frequency, $\Omega_{\rm K} = 1361$~Hz. The gravitational mass of the
  non-rotating star is $2.10$ $M_\odot$, which increases to $2.20$
  $M_\odot$ for rotation at $\Omega = \Omega_{\rm K}$.}
  \label{fig:profiles}
\end{figure}
\vspace {-12pt}

\begin{figure}[H]
\begin{center}
  \includegraphics[scale=0.4]{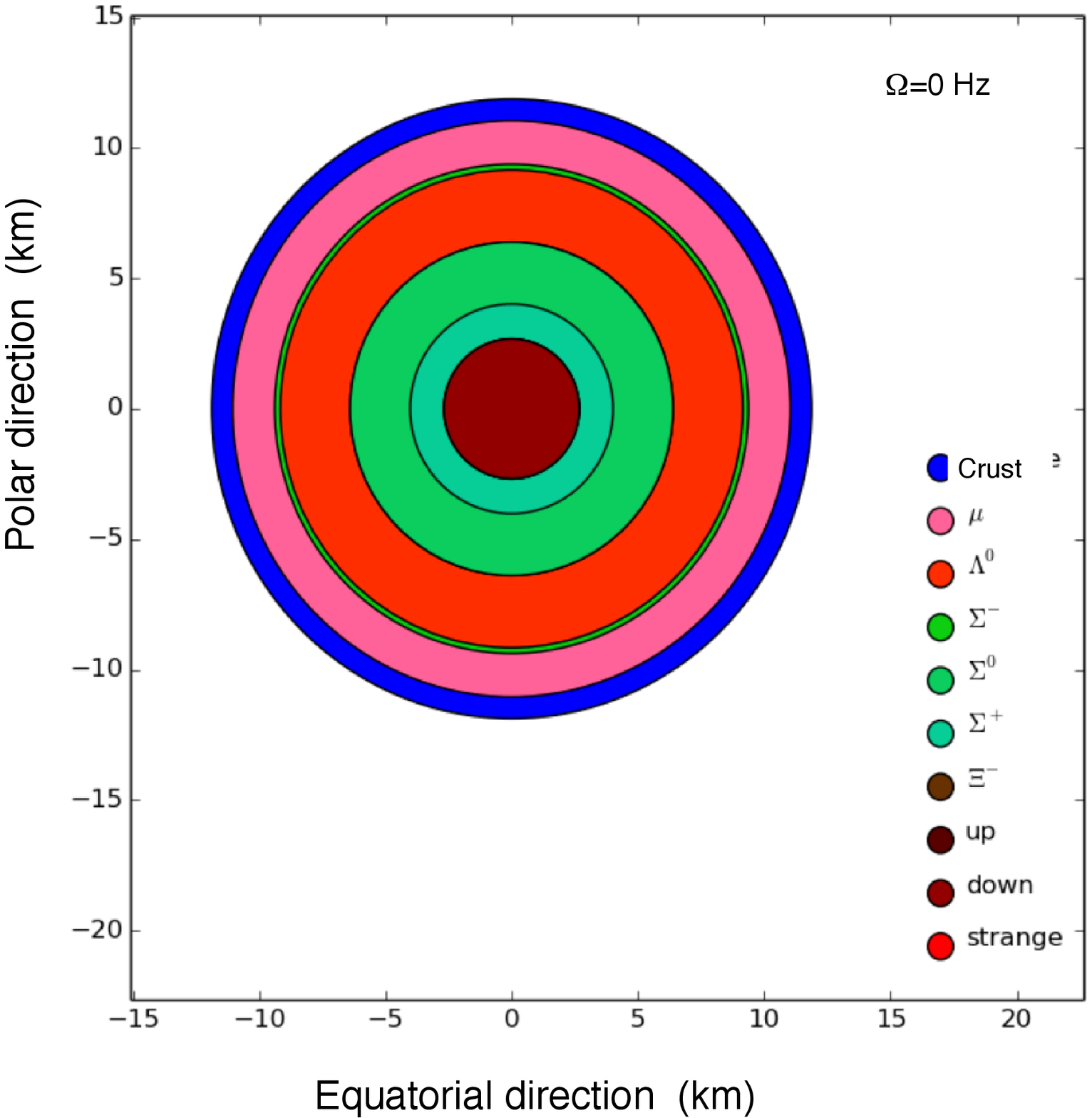}
      \includegraphics[scale=0.4]{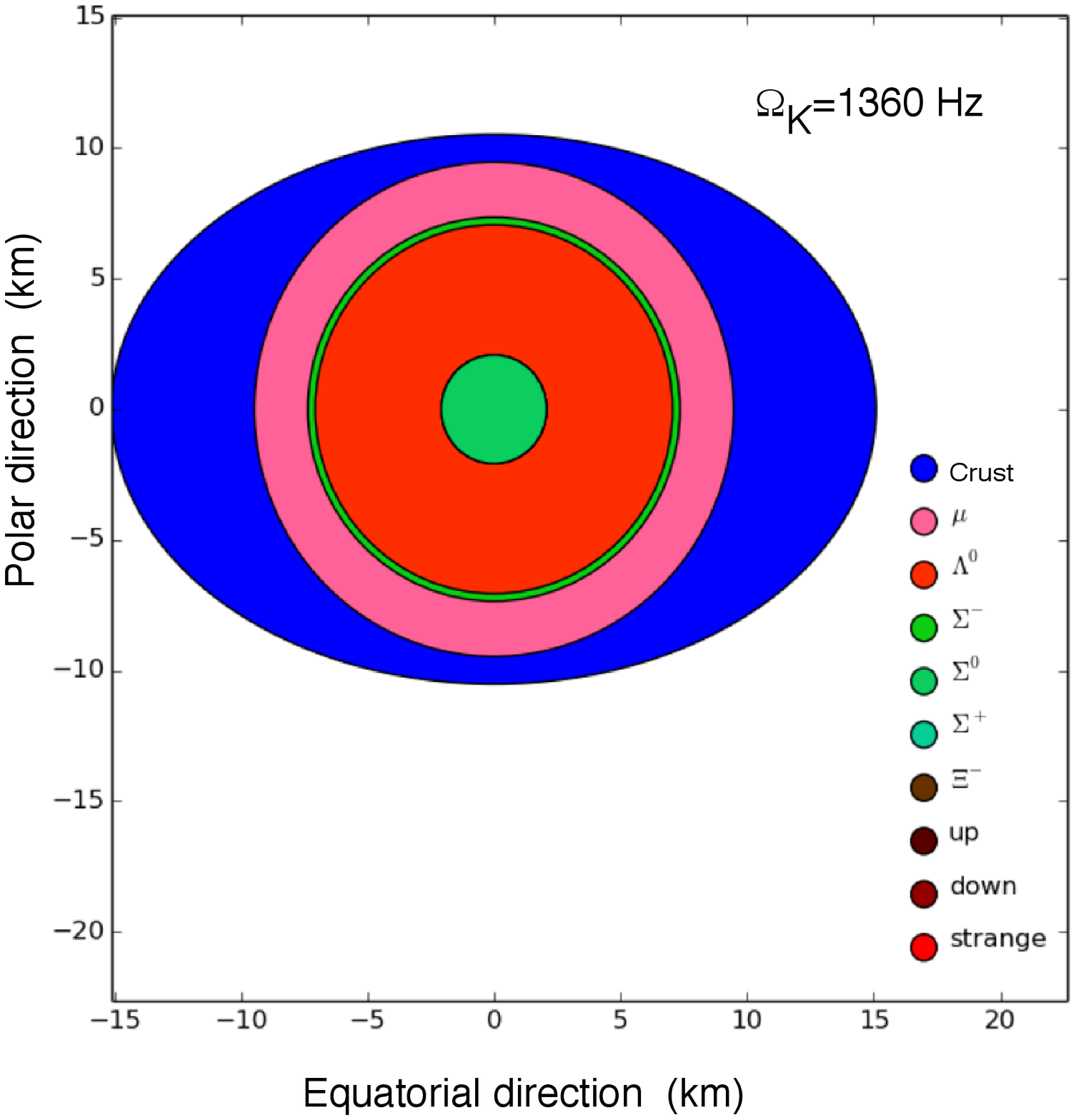}
\end{center}
\caption{Change of the interior composition of a $2$ $M_\odot$ neutron
  star caused by rotation, computed for the GM1 EoS. The vector
  interaction among quarks is $G_V=0.09 \, G_S$. The star on the
  left (right) hand-side is non-rotating (rotating at the Kepler
  frequency, $\Omega=\okgr$). The baryon number of both stars is the
  same ($\log_{10} A=57.51$).}
  \label{fig:comp1}
\end{figure}

A heat map showing the quark-hadron content of rotating neutron stars
computed for the GM1\linebreak ($G_V = 0.05 \, G_S$) equation of state is
shown in Figure\ \ref{GM1gv05QM}. As can be seen in this figure, up to
8\% of the total gravitational mass of these neutron stars exists in
the form a mixed quark-hadron phase. Lines of constant baryon number are also depicted on the figures as white
lines, labeled with the logarithm of the star's baryon number. These
lines were included to give a sense of the path that a secluded
neutron star would be expected to take through as it spins down.

\begin{figure}[H]
\begin{center}
\includegraphics[scale=0.8]{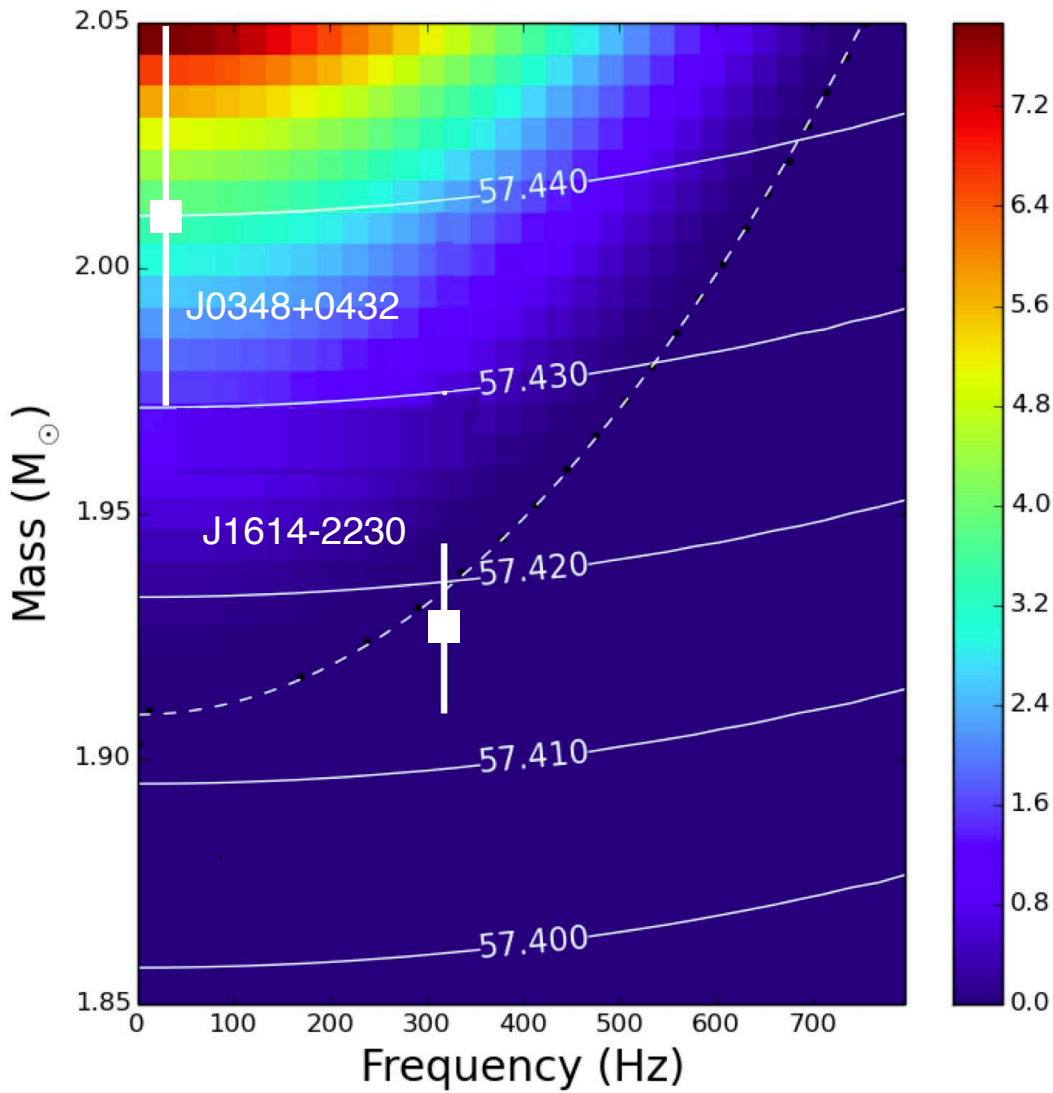}
\end{center}
\caption{Heat map showing the percent (column on the right) of total
  mass of a neutron star made up of deconfined quark matter, as
  predicted by the GM1 ($G_V =0.05 \, G_S$) EoS. The white solid lines
  show the rotational evolution of neutron stars with constant baryon
  numbers $A$ (the reported figures being $\log_{10} A$). Also shown
  are the observed masses of pulsars J1614--2230 and J0348+0432 
  and the
  trend line (dashed white) fit, which separates confined from
  deconfined matter. A fit of the trend line is given by Equation\
  (\ref{ThreshEqu}).}
\label{GM1gv05QM}
\end{figure}

By parsing out the maximum frequency where deconfined quark matter is
expected to exist at the center of the neutron star for a given mass
it is possible to get a curve through the gravitational
mass--frequency diagram for the threshold above which one can expect
to find deconfined quarks. These threshold frequencies for each
mass were fit to determine a (quadratic) function for the curve. The
fit equation, depicted as a dashed white line in Figure\ \ref{GM1gv05QM},
was found to have the form
\begin{equation}
\label{ThreshEqu} M(\Omega)  =  a \; \Omega^2 + c \, ,
\end{equation}
where $M$ is the neutron star's gravitational mass in solar masses,
$\Omega$ its  rotational frequency, and $a$ and
$c$ are parameters determined through fitting. The values for $a$ and
$c$ can be found in Table \ref{ThreshParams}. 

\begin{table}[H]
\caption{Parameters for the empirical deconfinement threshold curve
  for each equation of state (EoS) with the form shown in
  Equation\ (\ref{ThreshEqu}). }
\label{ThreshParams}
\centering
\begin{tabular}{c@{\hskip 1cm}c@{\hskip 1cm}c@{\hskip 1cm}}
\toprule
\textbf{EoS} & \textbf{a} (\boldmath $M_{\odot}~ s^2$) & \textbf{c} (\boldmath $M_{\odot}$)\\
\midrule
GM1~($G_V=0.05 \, G_S$)   & $2.48 \times 10^{-7}$ & $1.91$  \\
GM1~($G_V=0$)      & $2.75 \times 10^{-7}$ & $1.71 $ \\
DD2~($G_V=0$)      & $2.56 \times 10^{-7}$ & $1.89$ \\
\bottomrule
\end{tabular}
\end{table}
Figure \ref{GM1gv05QM} allows one to estimate the amount of
quark-hadron matter that may exist in the cores of neutron stars that
have both a measured frequency and mass, as illustrated for pulsars
PSR J1614--2230 ($M =1.928\pm 0.017 \, M_\odot$, rotational frequency
$f=318$~Hz) \cite{demorest10:a,fonseca16:a} and PSR J0348+0432
(\mbox{$M=2.01 \pm 0.04\, M_\odot$,} $f=26$~Hz)
\cite{lynch13:a,antoniadis13:a} in Figure \ref{GM1gv05QM}.  According
to this calculation, up to around 7\% of the mass of PSR J0348+0432
could be in the mixed quark-hadron phase, while the core of the more
rapidly rotating pulsars PSR J1614--2230 may be hovering right at the
quark deconfinement density.

\section{Discussion and Summary}\label{Sec:Discussion}

The type and structure of the matter in the cores of rotating neutron
stars (pulsars) depends on the spin frequencies of these stars
\cite{weber99:book,weber05:a,stejner09:a}, which opens up a new window
on the nature of matter deep in their cores.  We~find that, depending
on mass and rotational frequency, up to around 8\% of the mass of
massive neutron stars may be in the mixed quark-hadron phase, if the
quark-hadron phase transition is Gibbs-like. Examples of such stars
are pulsars PSR J1614--2230 with a gravitational mass of $1.928\pm
0.017\, M_{\odot}$ \cite{fonseca16:a} and PSR J0348+0432 with a~mass
of $2.01 \pm 0.04 \, M_\odot$ \cite{lynch13:a,antoniadis13:a}
(Figure~\ \ref{GM1gv05QM}).  Pure quark matter in the centers of neutron
stars is not obtained for any of the models for the nuclear equation
of state studied in this work.  We also find that the gravitational
mass at which quark deconfinement occurs in rotating neutron stars
varies quadratically with spin frequency, which~can be fitted by a~simple quadratic formula.

Our view of the interior composition of pulsars has changed
dramatically since their first discovery almost 50 years ago. It has
also become clear during that time period that all the ambient
conditions that characterize pulsars tend to the extreme, making
pulsars almost ideal astrophysical laboratories for a broad range of
physical studies. Owing to the unprecedented wealth of high-quality
data on pulsars provided by radio telescopes, X-ray~satellites--and
soon the latest generation of gravitational-wave detectors--it seems
within reach to decipher the inner workings of pulsars, and to explore
the phase diagram of cold and ultra-dense hadronic matter from
astrophysics.



\vspace{6pt} 

\supplementary{Several movies showing the changing particle
  compositions inside of rotating neutron stars are available online
  at http://www-rohan.sdsu.edu/\!$\sim$fweber/Mellinger/Compositions.html.}

\acknowledgments{This work is supported through the U.S. National
  Science Foundation under Grants PHY-1411708 and DUE-1259951.
  Additional computing resources are provided by the Computational
  Science Research Center and the Department of Physics at San Diego
  State University.  Milva G. Orsaria thanks the financial support from
  American Physical Society's International Research Travel Award
  Program.  Gustavo~A.~Contrera, and Milva G. Orsaria acknowledge financial
  support from CONICET and UNLP (Project identification code 11/G140
  and 11/X718), Argentina.}
 
\authorcontributions{The authors contributed equally to the
  theoretical and numerical aspects of the work presented in this
  paper. } 

\conflictofinterests{The authors declare no conflict of interest.}





\renewcommand\bibname{References}

\end{document}